\documentclass[%
 preprint,
superscriptaddress,
 amsmath,amssymb,
 aps,
]{revtex4-1}

\usepackage{graphicx}
\usepackage{dcolumn}
\usepackage{bm}
\usepackage{amsfonts}
\usepackage{hyperref}
\usepackage[mathlines]{lineno}

\newcommand{\snn}{\sqrt{s_\text{NN}}}

\begin{document}


\title{Study of neutron density fluctuation and neutron-proton correlation in Au+Au collisions using PYTHIA8/Angantyr}

\author{Zuman Zhang}\thanks{Email: zuman.zhang@hue.edu.cn}
\affiliation{School of Physics and Mechanical Electrical \& Engineering, Hubei University of Education, Wuhan 430205, China}
 \affiliation{Institute of Theoretical Physics, Hubei University of Education, Wuhan 430205, China}
 \affiliation{Key Laboratory of Quark and Lepton Physics (MOE), Central China Normal University, Wuhan 430079, China}
 \author{Sha Li}\thanks{Email: lisha@hue.edu.cn}
 \affiliation{School of Physics and Mechanical Electrical \& Engineering, Hubei University of Education, Wuhan 430205, China}
\author{Ning Yu}\thanks{Email: ning.yuchina@gmail.com}
\affiliation{School of Physics and Mechanical Electrical \& Engineering, Hubei University of Education, Wuhan 430205, China}
 \affiliation{Institute of Theoretical Physics, Hubei University of Education, Wuhan 430205, China}
 \affiliation{Key Laboratory of Quark and Lepton Physics (MOE), Central China Normal University, Wuhan 430079, China}
 \author{Jianping Lin}\thanks{Email: jianpinglin0905@163.com}
 \affiliation{School of Physics and Mechanical Electrical \& Engineering, Hubei University of Education, Wuhan 430205, China}
 \author{Shuang Li}\thanks{Email: lish@ctgu.edu.cn}
 \affiliation{College of Science, China Three Gorges University, Yichang 443002, China}
 \affiliation{Center for Astronomy and Space Sciences, China Three Gorges University, Yichang 443002, China}
 \author{Siyu Tang}\thanks{Email: tsy@wtu.edu.cn}
 \affiliation{School of Mathematical \& Physical Sciences, Wuhan Textile University, Wuhan 430200, China}
\author{Daicui Zhou}\thanks{Email: dczhou@mail.ccnu.edu.cn}
 \affiliation{Key Laboratory of Quark and Lepton Physics (MOE), Central China Normal University, Wuhan 430079, China}

\date{\today}

\begin{abstract}
Utilizing the PYTHIA8 Angantyr model, which incorporates the  multiple-parton interactions (MPI) based color reconnection (CR) mechanism, we study the relative neutron density fluctuation and neutron-proton correlation in Au+Au collisions at  $\snn$ = 7.7, 11.5, 14.5, 19.6, 27, 39, 62.4, and 200 GeV. In this study, we have not only delved into the dependence of these two remarkable observations on rapidity, centrality, and energy, but also presented an analysis of their interplay with the MPI and CR. Our results have shown that the light nuclei yield ratio of proton, deuteron, and triton, expressed by the elegant expression $N_tN_p/N_d^2$, remains unchanged even as the rapidity coverage and collision centrality increase. Interestingly, we have also revealed that the effect of CR is entirely dependent on the presence of MPI; CR has no  impact on the yield ratio if MPI is off. Our findings further demonstrate that the light nuclei yield ratio experiences a slight increase with increasing collision energy as predicted by the PYTHIA8 Angantyr model, but it cannot describe the non-monotonic trend observed by the STAR experiment. Based on the Angantyr model simulation results, it is essential not to overlook the correlation between neutron and proton fluctuations. The Angantyr model is a good baseline for studying collisions in the absence of a Quark-Gluon Plasma (QGP) system, given its lack of flow and jet quenching.

\begin{description}\item[PACS numbers]
\verb+ relativistic heavy-ion collisions, quark deconfinement,+quark-gluon plasma production, phase transitions
\end{description}
\end{abstract}

\maketitle

\section{Introduction}
\label{sec:intro}

  The creation of a state of matter known as the quark-gluon plasma (QGP), a mixture of deconfined quarks and gluons, is believed to occur in heavy-ion collisions at ultra-relativistic energies at extremely high temperatures and/or densities. Understanding the Quantum Chromodynamics (QCD) phase diagram of strongly interacting matter is crucial in the field of nuclear physics. The QCD phase diagram can be represented in a two-dimensional graph of temperature (T) versus baryon chemical potential ($\mu_{\mathrm{B}}$). According to lattice QCD calculations, the transition from the hadronic phase to the QGP occurs as a smooth crossover at low values of $\mu_{\mathrm{B}}$~\cite{ref:QCDphaseplot,Aoki:2006we}. However, at finite values of $\mu_{\mathrm{B}}$, it is predicted to be a first-order phase transition based on QCD model calculations~\cite{Endrodi:2011gv,ref:LQCDCplot,ref:lqcd03,ref:lqcd05}. If these predictions hold true, there must exist a QCD critical point, marking the endpoint of the first-order phase boundary. Despite ongoing theoretical discussions on the location and even the existence of the QCD critical point, relativistic heavy-ion collisions provide a controlled means to explore the QCD phase structure and search for the critical point~\cite{ref:SPS_result,ref:RHIC_result,Gupta:2011wh,Luo:2017faz,Bzdak:2019pkr,Aggarwal:2010wy,Luo:2015doi,Adamczyk:2014fia,Adamczyk:2013dal,Adamczyk:2017wsl,Adam:2020unf}.

  As the system approaches the QCD critical point, the correlation length and density fluctuations experience an increase. The fluctuations of conserved quantities, such as net-baryon, net-charge, and net-strangeness, are acutely influenced by the correlation length. The STAR experiment has conducted measurements of high-order cumulants and second-order off-diagonal cumulants of net-proton, net-charge, and net-kaon multiplicity distributions~\cite{RN182,RN183,RN184,RN185} in Au+Au collisions across a broad range of energies, from as low as $\snn$ = 7.7 GeV up to 200 GeV in the most central collisions, a non-monotonic behavior of the fourth-order net-proton cumulant ratio was observed, with a minimum around 19.6 GeV.
  The non-monotonic behavior observed in the fourth-order net-proton cumulant ratio in the STAR experiment cannot be explained by existing model calculations unless the physics of the QCD critical point is considered.

  Moreover, as the correlation length increases and instability spinodal domains form, critical fluctuations and first-order phase transitions can give rise to significant baryon density fluctuations. It is predicted that the production of light nuclei is sensitive to the baryon density fluctuations, and thus can be used to probe the QCD phase transition in heavy-ion collisions~\cite{Sun:2017xrx,Deng:2020zxo,Shuryak:2019ikv,Yu:2018kvh,Shao:2019xpj,RN9}. For instance, the light nuclei yield ratio $N_pN_t/N_d^2$ of produced proton ($p$), deuteron ($d$), and triton ($t$) can be described by the relative neutron density fluctuation $\langle (\delta n)^2\rangle /\langle n\rangle^2$ (also denoted as $\Delta n$ in Ref.~\cite{Sun:2017xrx}). Interestingly, it was observed that the yield ratio and the extracted neutron density fluctuation in central Au+Au collisions measured by STAR experiment show clear non-monotonic energy dependence, with a peak at $\snn$ = $20\sim30$ GeV~\cite{Zhang:2019wun,Zhang:2020ewj}.

  In this paper, we studied the light nuclei yield ratio in Au+Au collisions at $\sqrt{s_{\mathrm{NN}}}$ = 7.7, 11.5, 14.5, 19.6, 27, 39, 62.4, and 200 GeV from the framework of PYTHIA8 Angantyr model. Angantyr adds a framework wherein AA-collisions can be constructed as a superposition of binary nucleon-nucleon collisions as implemented in the PYTHIA8 event generator.
  Our paper is organized as following: In Sec.~\ref{subsec:Angantyr}, we give a brief introduction to the PYTHIA8 Angantyr model. In Sec.~\ref{sec:2correl}, the relationship between the relative neutron density fluctuation and the light nuclei yield ratio in heavy ion collisions is given. In Sec.~\ref{sec:result}, we present the behavior of neutron density fluctuation and neutron-proton correlation ($\alpha$) . Finally, the summary will be given in Sec.~\ref{sec:summary}.

\section{Event generation and definition of light nuclei yield ratio}
\label{sec:Generation_Methodology}
\subsection{PYTHIA8 (Angantyr) model}
\label{subsec:Angantyr}
  PYTHIA~\cite{Sjostrand:2006za} is an event generator that is extensively and successfully used for the study of proton-proton and proton-lepton collisions. In pp collisions, Multi-Parton Interaction (MPI) is generated under the assumption that every partonic interaction is almost independent. PYTHIA8 natively does not support heavy-ion systems.
  Recently, PYTHIA8 Angantyr model~\cite{Bierlich:2018xfw} extrapolates the pp dynamics into heavy-ion collisions using the PYTHIA8 event generator, enables the study of heavy nuclei collisions, namely proton-nuclei (pA) and  nuclei-nuclei (AA).
  The Angantyr model combines several nucleon-nucleon collisions into one heavy-ion collision. It is a combination of many-body physics (theoretical) models suitable for producing hard and soft interactions, initial and final-state parton showers, particle fragmentations, multi-partonic interactions, color reconnection (CR) mechanisms, and decay topologies. But it doesn't include any mechanism of QGP medium believed to be produced in AA collisions.

  In the current version of the PYTHIA8 Angantyr model~\cite{Sjostrand:2008za}, in a heavy-ion collision, each projectile nucleon can interact with several target nucleons, and the number of participant nucleons determined by Glauber model. This model added several algorithms to distinguish different types of nucleon-nucleon interactions, such as elastic, diffractive, and absorptive. It's supposed to well describe the final-state properties, such as multiplicity and transverse momentum distributions in AA collisions~\cite{Abdel:2022,Samsul:2022}.

  We use 8.308 version of PYTHIA8 in this work. The simulation is considered with different PYTHIA tunes using MPI and CR configurations. About one million events for each collision energy were generated for the Au + Au collisions. In the PYTHIA8 Angantyr model, due to the both density and density fluctuation are dependent on system volume. To remove the system volume effect, we use two dimensionless statistical quantities $\langle (\delta p)\rangle /\langle p\rangle$  and $\langle (\delta n)\rangle /\langle n\rangle$ (Sec.~\ref{sec:2correl}).
  The nucleons are extracted for different rapidity ranges and centralities. We define the centrality intervals which are based on the summed transverse energy ($\sum E_T$) in the pseudorapidity interval [-0.5, 0.5].

\subsection{Definition of light nuclei yield ratio}
\label{sec:2correl}

  Based on the references cited as~\cite{RN9,RN172,RN2022}, it has been proposed that the creation of light nuclei occurs through the process of nucleon coalescence. By this mechanism, protons and neutrons come together to form deuterons and tritons. If we neglect the binding energy,
  the number of these particles can be expressed as

  \begin{equation}
  N_d \approx
  \frac{3}{2^{1/2}}\left(\frac{2\pi}{mT}\right)^{3/2}N_p\langle n \rangle(1+\alpha)
  \label{eq:N_d}
  \end{equation}

  \begin{equation}
  N_t \approx
  \frac{3^{3/2}}{4}\left(\frac{2\pi}{mT}\right)^{3}N_p\langle n \rangle^2(1+\Delta n+2\alpha),
  \label{eq:N_t}
  \end{equation}
  then, to eliminate the energy dependence on the local effective temperature ($T$) at coalescence, we have the light nuclei ratio
  \begin{equation}
    \frac{N_tN_p}{N_d^2}=g\frac{1+\Delta n+2\alpha}{(1+\alpha)^2},
    \label{lnr1}
  \end{equation}
  where $N_p$, $N_d$,  and $N_t$ represent the number of protons, neutrons, and tritons respectively. $m$ stands for the mass of a nucleon. $\Delta n = \langle (\delta n)^2\rangle /\langle n\rangle^2$ is the dimensionless relative neutron density fluctuation. The standard deviation and mean value of neutron production are represented by $\langle (\delta n) \rangle$ and $\langle n\rangle$ in our work. $\alpha= \langle\delta n \delta p\rangle/(\langle n\rangle\langle p\rangle)$ is the neutron and proton number density correlation coefficient. $g$ = $\frac{3^{3/2}}{4}/(\frac{3}{2^{1/2}})^{2}$ = $\frac{1}{2\sqrt{3}}$.

   If we assume correlation coefficient between neutron and proton density is small, then $\alpha$ can be approximated as zero ($\alpha\approx0$), Eq.~(\ref{lnr1}) can be rewritten as
  \begin{equation}
    \frac{N_tN_p}{N_d^2}=\frac{1+\Delta n}{2\sqrt{3}}.
    \label{lnr2}
  \end{equation}

\section{Results and Discussions}
\label{sec:result}

  \begin{figure}[tb]
    \includegraphics[width=0.47\textwidth]{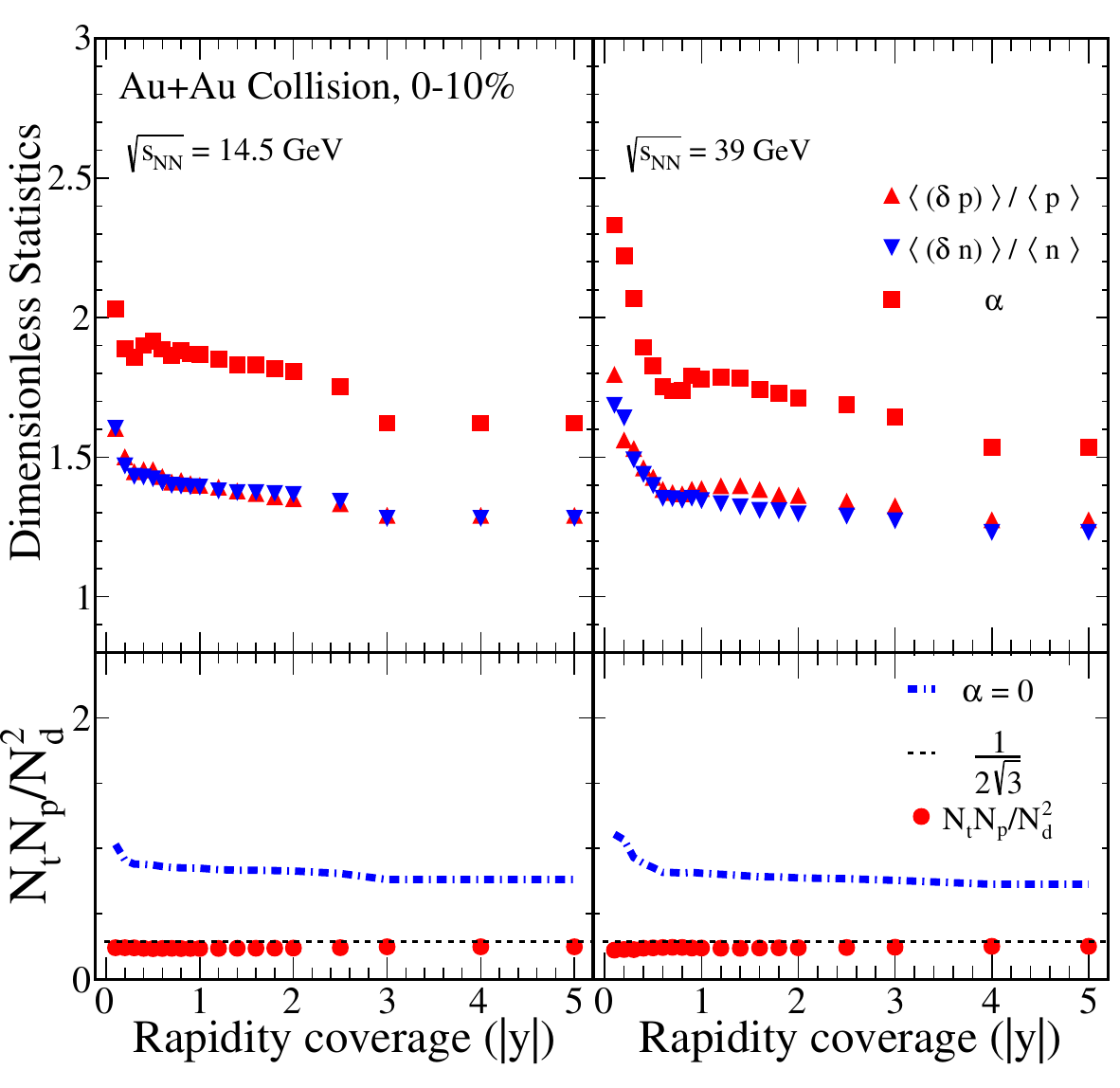}
    \caption{\label{fig1}
    In the top panels of the figure, various dimensionless statistics are displayed, including $\langle (\delta p) \rangle /\langle p \rangle$, $\langle (\delta n) \rangle /\langle n \rangle$, and $\alpha$, which are associated with 0-10\% Au+Au collisions at center-of-mass energies of $\snn=$ 14.5 and 39 GeV . Meanwhile, the bottom panels depict the ratio of yields for light nuclei, $N_tN_p/N_d^2$, which is calculated using the data from the top panels, represented as solid circles in accordance with Eq.~(\ref{lnr1}), and as a dash-dot line in accordance with Eq.~(\ref{lnr2}).
    }
  \end{figure}

  Except for the physical parameters utilized in the model calculations mentioned in last paragraph of Sec.~\ref{subsec:Angantyr}, we have the option to incorporate the multiple-parton interactions (MPI) based color reconnection (CR) mechanism by switching on/off the ColourReconnection:reconnect
  and PartonLevel:MPI.

\begin{figure}[tb]
    \includegraphics[width=0.47\textwidth]{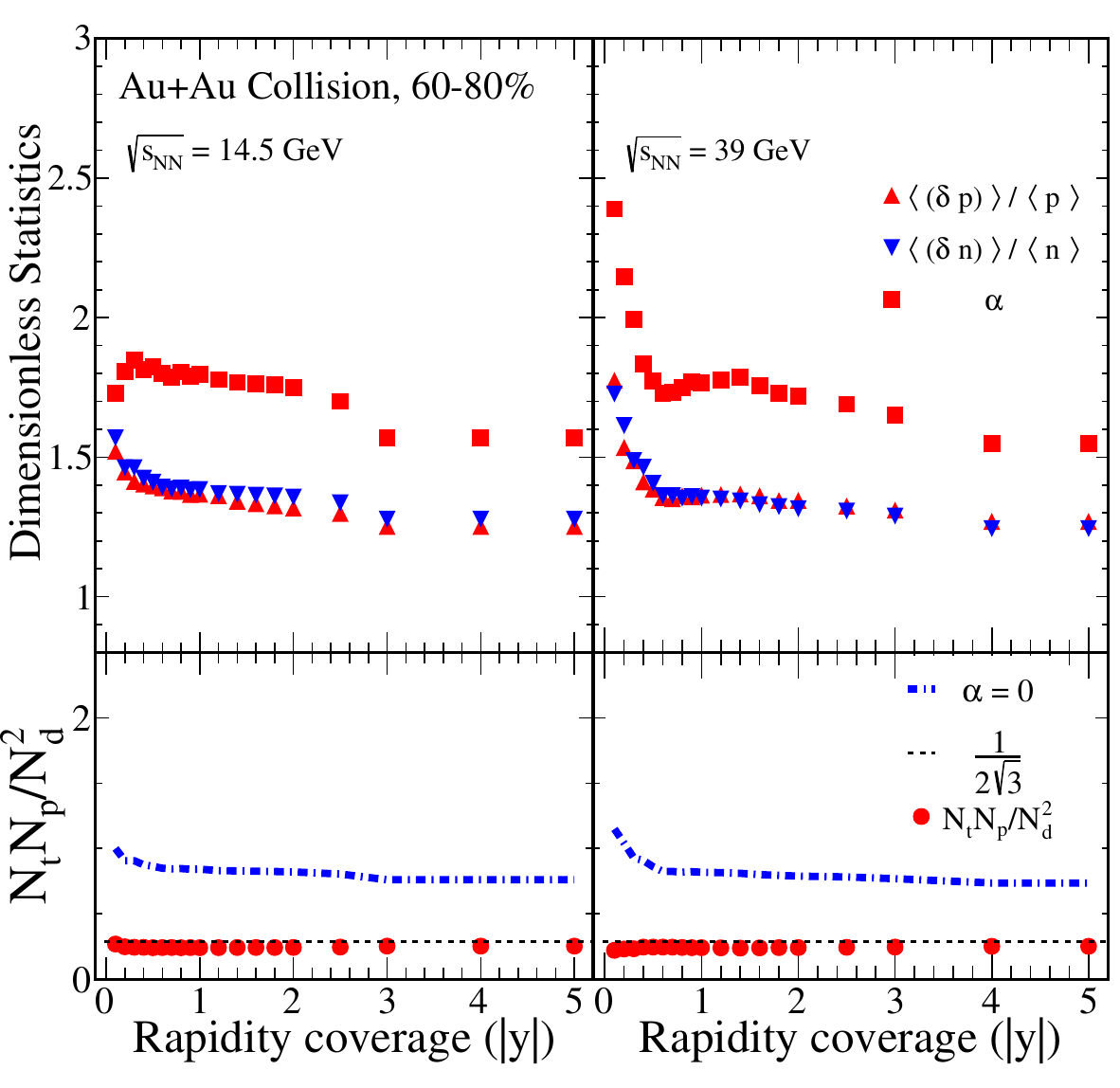}
    \caption{\label{fig2}
    In the top panels of the figure, various dimensionless statistics are displayed, including $\langle (\delta p)\rangle /\langle p\rangle$, $\langle (\delta n)\rangle /\langle n\rangle$ and $\alpha$, which are associated with 60-80\% Au+Au collisions at center-of-mass energies of $\snn=$ 14.5 and 39 GeV . Meanwhile, the bottom panels depict the ratio of yields for light nuclei, $N_tN_p/N_d^2$, which is calculated using the data from the top panels, represented as solid circles in accordance with Eq.~(\ref{lnr1}), and as a dash-dot line in accordance with Eq.~(\ref{lnr2}).}
\end{figure}

\begin{figure}[tb]
    \includegraphics[width=0.45\textwidth]{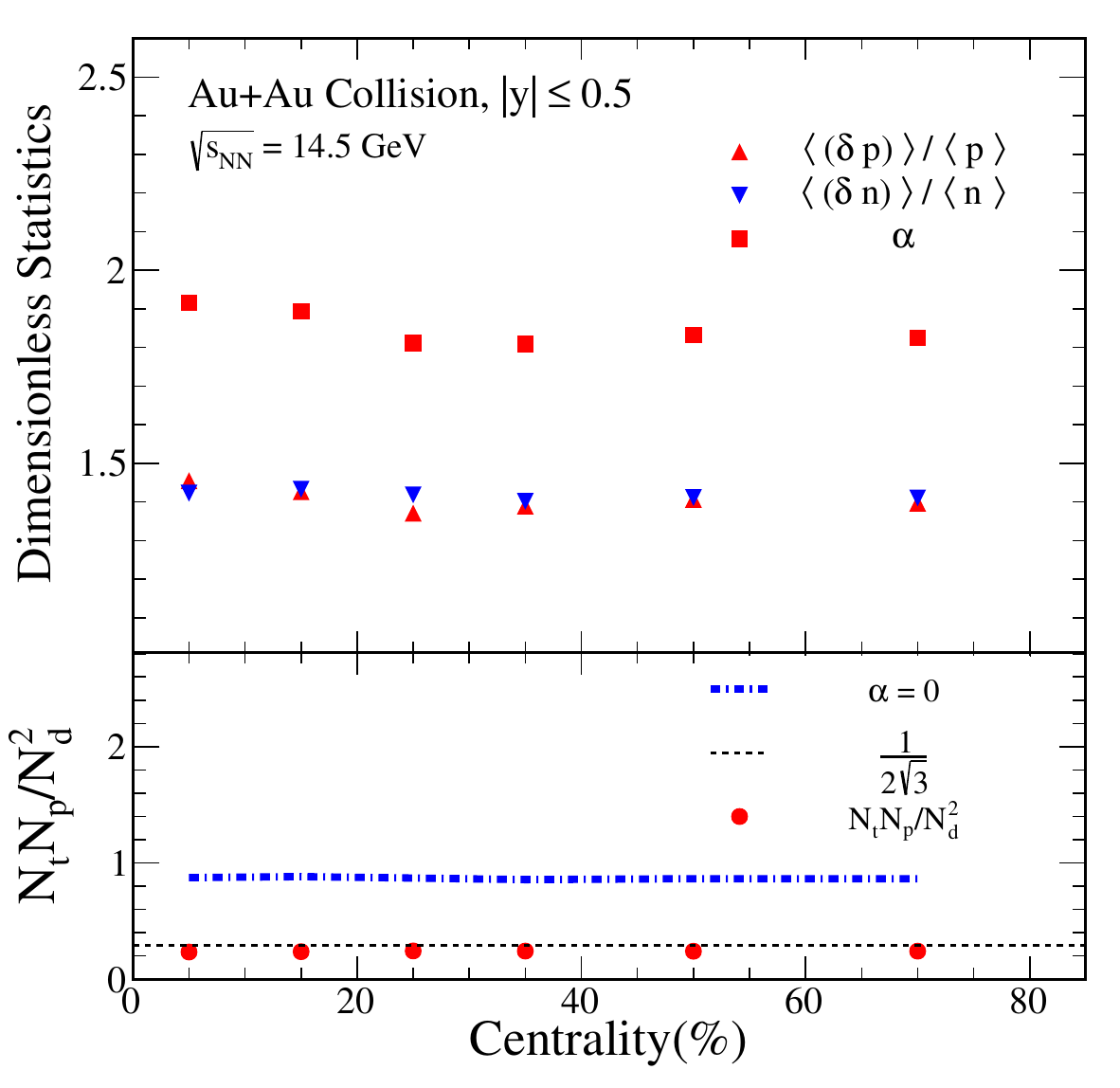}
    \includegraphics[width=0.45\textwidth]{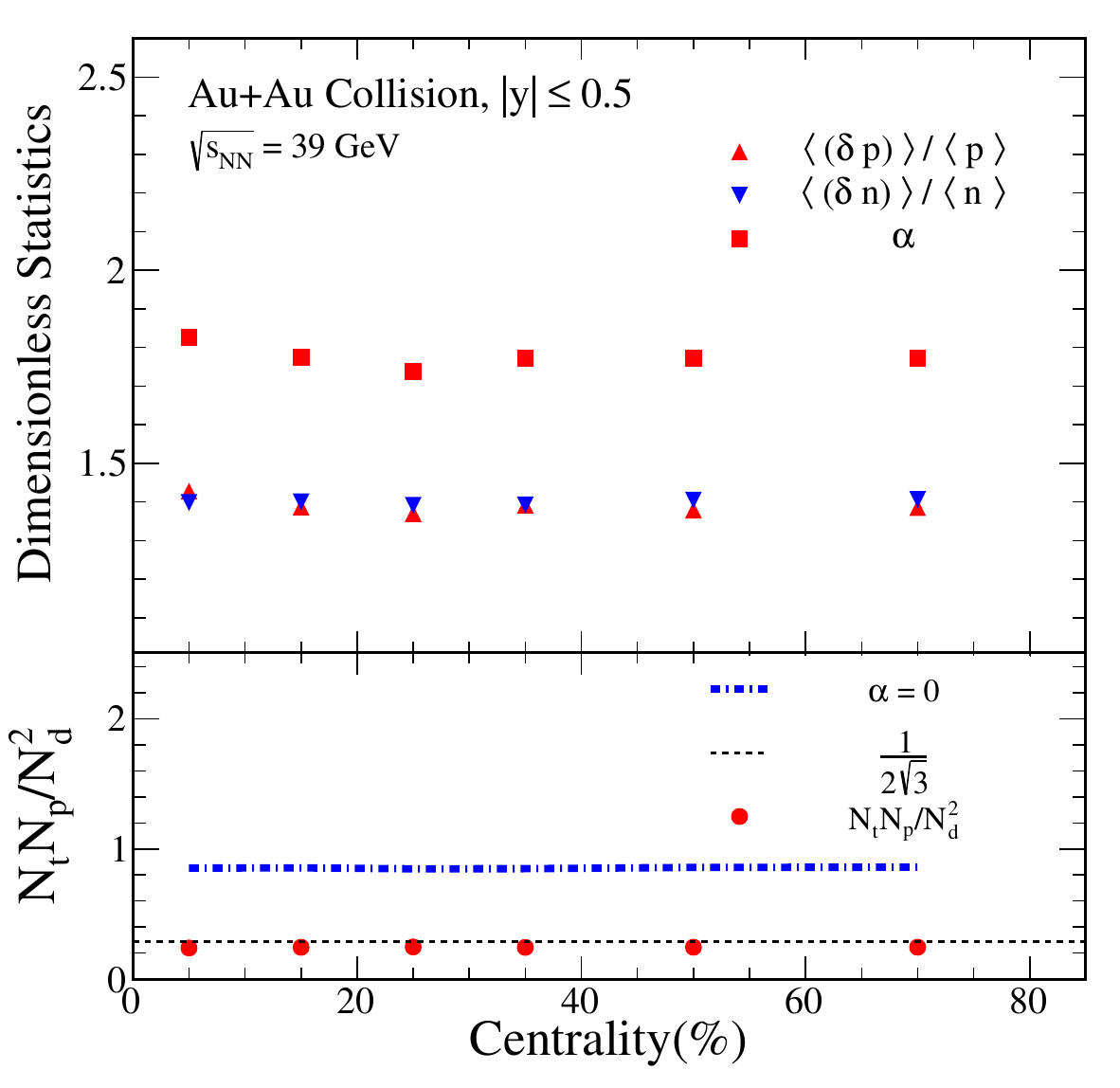}
    \caption{\label{fig3} The dimensionless statistics $\langle (\delta p)\rangle /\langle p\rangle$, $\langle (\delta n)\rangle /\langle n\rangle$ and $\alpha$ for Au+Au collisions at center-of-mass energies of $\snn=$ 14.5 and 39 GeV, with rapidities confined to the range $|y|\leq 0.5$. In addition, the ratio of yields for light nuclei, $N_tN_p/N_d^2$, is shown in the same figure, represented as solid circles in accordance with Eq.~(\ref{lnr1}), and as a dash-dot line in accordance with Eq.~(\ref{lnr2}).
    }
\end{figure}

\begin{figure}[tb]
    \includegraphics[width=0.49\textwidth]{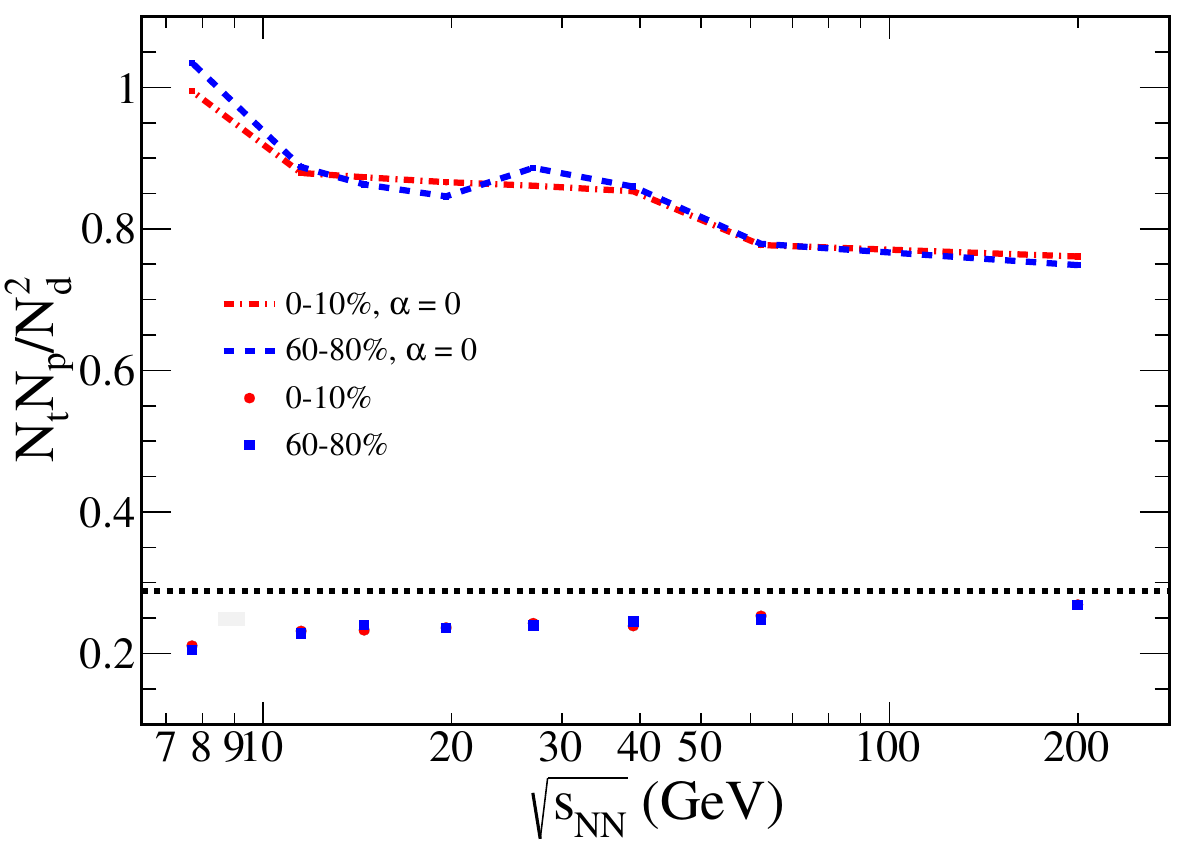}
    \caption{\label{fig4}
    The PYTHIA8 Angantyr model was used to examine the collision energy dependence of the light nuclei yield ratio $N_tN_p/N_d^2$ in Au+Au collisions with $|y|\leq 0.5$. The results revealed solid circles and squares for 0-10\% central and 60-80\% peripheral collisions, respectively. Additionally, the dash-dot lines displayed the corresponding results with vanished $\alpha$.
    }
\end{figure}

  The results in Figure~\ref{fig1},~\ref{fig2},~\ref{fig3},~\ref{fig4}, and~\ref{fig6} have been investigated using the PYTHIA8 Angantyr model with the multiple-parton interactions based color
  reconnection mechanism.

  The top panels depicted in Figure~\ref{fig1} illustrate
  the rapidity dependence of the dimensionless statistics $\langle (\delta p)\rangle /\langle p\rangle$ and $\langle (\delta n)\rangle /\langle n\rangle$ for 0-10$\%$ Au+Au collisions at $\snn=$ 14.5 and 39 GeV.
  As the rapidity coverage is increased, the relative fluctuations in nucleon density ($\langle (\delta p)\rangle /\langle p\rangle$  and $\langle (\delta n)\rangle /\langle n\rangle$) displayed in the top panels of Figure~\ref{fig1} appear to decrease.
  As the rapidity coverage is expanded, the nucleon density fluctuations appear to converge towards a constant value. The $\langle (\delta p)\rangle /\langle p\rangle$  and $\langle (\delta n)\rangle /\langle n\rangle$ are consistent in Figure~\ref{fig1}.  The correlation $\alpha$ have similar trend as relative fluctuations in nucleon density with rapidity coverage  for 0-10$\%$ Au+Au collisions at both $\snn=$ 14.5 and 39 GeV.
  We plot the line of $1/2\sqrt{3}$, which means both density fluctuation and correlation disappeared.
  From Eq.~(\ref{lnr1}), the light nuclei ratio $N_tN_p/N_d^2$ can be calculated. The rapidity dependence of this ratio, the solid circles, is shown in the bottom panel of Figure~\ref{fig1}.
  Both the ratio for 0-10$\%$ Au+Au collisions at $\snn=$ 14.5 and 39 GeV are lower than the line of $1/2\sqrt{3}$ with rapidity distribution. One the other hand, the light nuclei ratio with the dash-dot lines ($\alpha\approx0$) higher than the line of $1/2\sqrt{3}$. So, the $\alpha$ can not be neglected in the central collision to calculate the relative neutron density fluctuation from the light nuclei yield ratio with Eq.~(\ref{lnr2}). Results from other collision energies at central collision are similar.

   The rapidity dependence of $\langle (\delta p)\rangle /\langle p\rangle$  and $\langle (\delta n)\rangle /\langle n\rangle$ for 60-80$\%$ Au+Au collisions at $\snn=$ 14.5 and 39 GeV in the top panels, and related light nuclei ratio in the bottom panels are shown in Figure~\ref{fig2}. Similar to the results of the central collisions, $\langle (\delta p)\rangle /\langle p\rangle$  and $\langle (\delta n)\rangle /\langle n\rangle$ decrease with the rapidity coverage is increased. The correlation $\alpha$ is dependent on rapidity coverage and has similar trend as relative fluctuations in nucleon density.
   We present the light nuclei yield ratios, calculated using  Eq.~(\ref{lnr1}) and Eq.~(\ref{lnr2}), in the bottom lower section of Figure~\ref{fig2}. Our results reveal the exclusion of the $\alpha$ parameter leads to a significant increase in the light nuclei yield ratio.
   In other words, the omission of $\alpha$ results in a smaller relative neutron density fluctuation with the light nuclei yield ratio.

   The top panels of Figure~\ref{fig3} show the centrality dependent $\langle (\delta p)\rangle /\langle p\rangle$  and $\langle (\delta n)\rangle /\langle n\rangle$ for Au+Au collisions at $\snn=$ 14.5 and 39 GeV with rapidity coverage of $|y|\leq 0.5$. All two quantities are flat from central to peripheral collision.  At the bottom of Figure~\ref{fig3}, it can be found that the two ratios given by Eq.~(\ref{lnr1}) and Eq.~(\ref{lnr2}) are flat at central, mid-central, and peripheral collisions. It is noted that the relative nucleon density fluctuation cannot be extracted directly from the light nuclei ratio, and the effects of neutron-proton correlation $\alpha$ must be considered in different centralities of collisions.

   The collision energy dependence of the  $N_tN_p/N_d^2$ from 0-10\% central to 60-80\% peripheral Au+Au collisions with $|y|\leq 0.5$ are shown in Figure~\ref{fig4}.  It is clear from the Figure~\ref{fig4} that the light nuclei yield ratio increases slightly with increasing collision energy from the PYTHIA8 Angantyr model.  The light nuclei yield ratio of peripheral collisions is similar to that of central collisions, and both results are lower than $1/2\sqrt{3}$. The results of vanished $\alpha$ decreases with increasing collision energy are also shown as dash-dot lines. In Au+Au collisions, both in central (0-10\%) and peripheral (60-80\%) collisions, the yield ratio of light nuclei is not highly consistent with each other after the disappearance of
   $\alpha$.

\begin{figure}[tb]
    \includegraphics[width=0.49\textwidth]{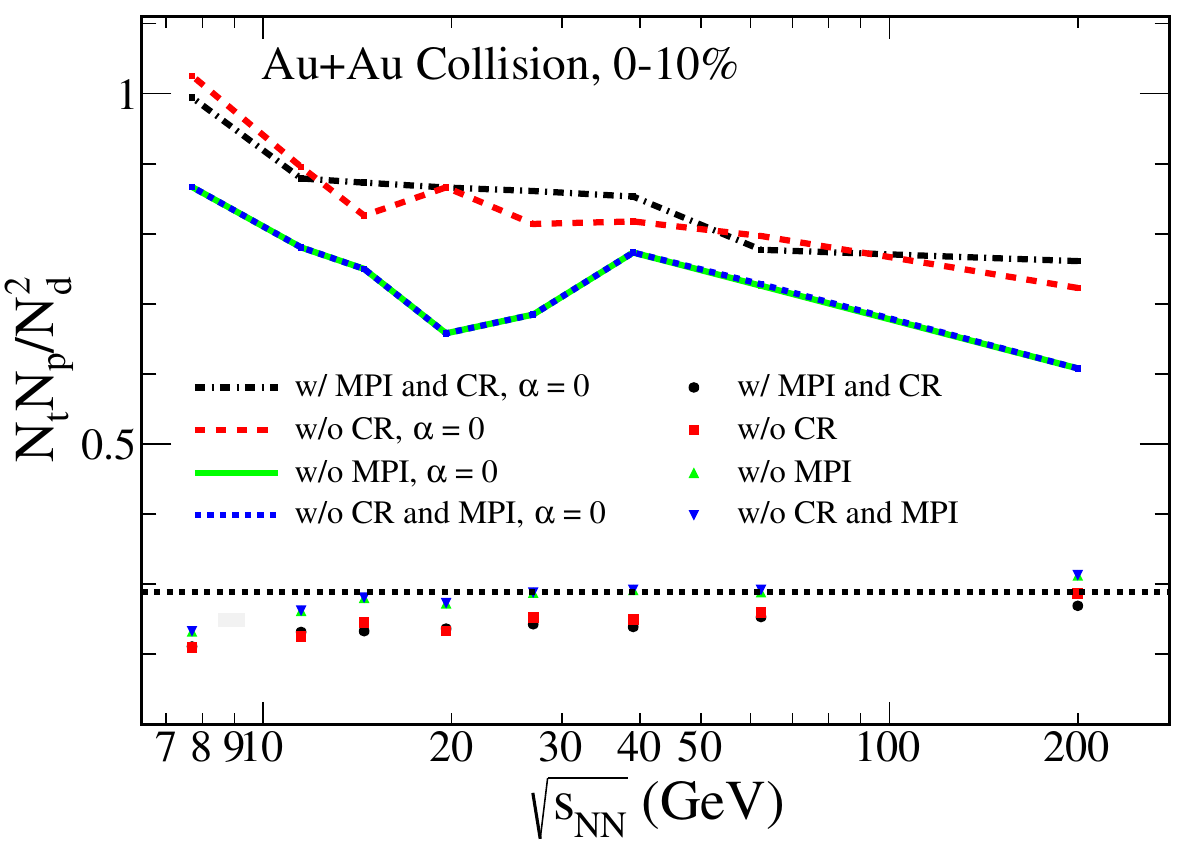}
    \includegraphics[width=0.49\textwidth]{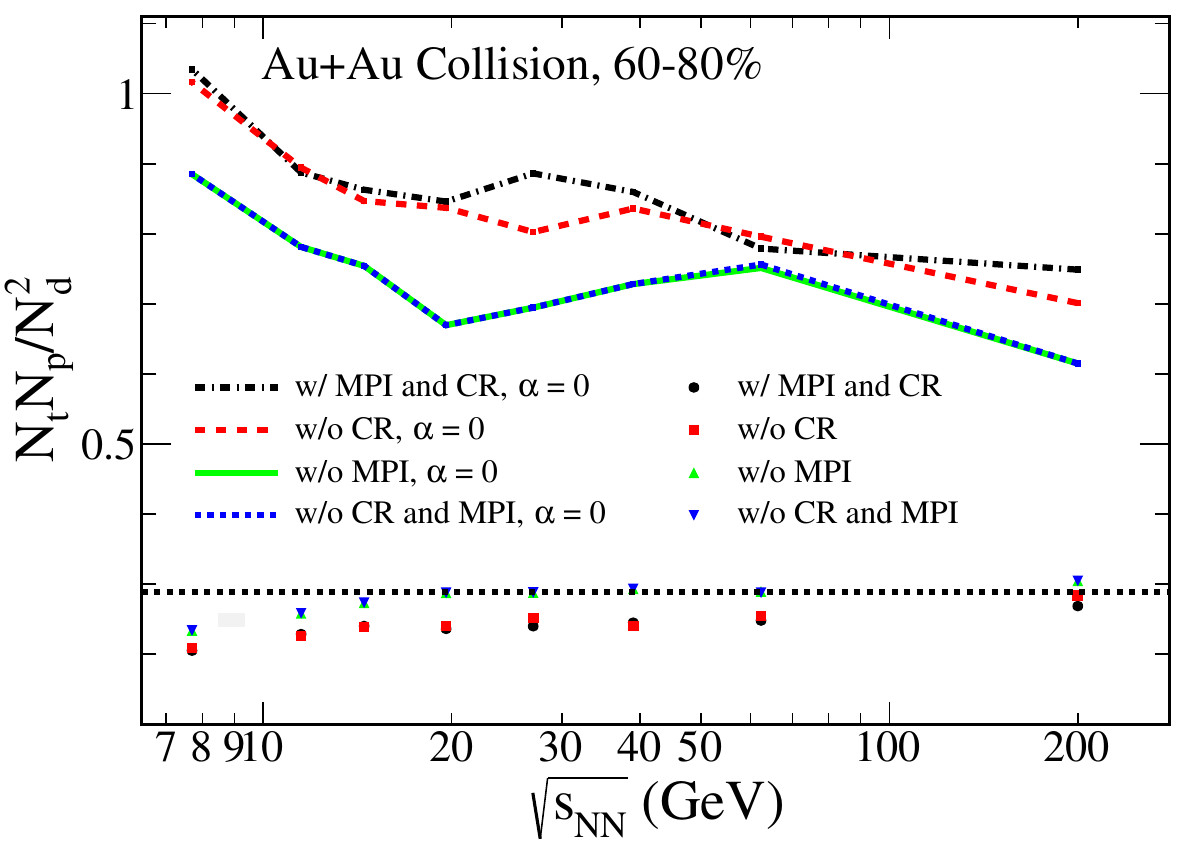}
    \caption{\label{fig5}
    The PYTHIA8 Angantyr model was used to examine the collision energy dependence of the light nuclei yield ratio $N_tN_p/N_d^2$ in Au+Au collisions with $|y|\leq 0.5$ in different PYTHIA8 Angantyr model tunes. The results shown in upper and lower panel related to 0-10\% central and 60-80\% peripheral collisions, respectively.
    }
\end{figure}

  To see the effect of different PYTHIA tunes, we consider the following configurations: MPI with CR, No CR, No MPI, and both MPI and CR off. Figure~\ref{fig5} shows the collision energy dependence of the light nuclei yield ratio $N_tN_p/N_d^2$ from 0-10\% central and 60-80\% peripheral Au+Au collisions with $|y|\leq 0.5$ in different PYTHIA8 Angantyr model tunes.  It is  also clear from this figure that the light nuclei yield ratio increases slightly with increasing collision energy from the PYTHIA8 Angantyr model. But the results of vanished $\alpha$ decreases with increasing collision energy are also shown as dash-dot lines. At Au+Au collisions, in different PYTHIA8 Angantyr model tunes, there is no effect of CR if MPI is off.

 \begin{figure}[tb]
    \includegraphics[width=0.49\textwidth]{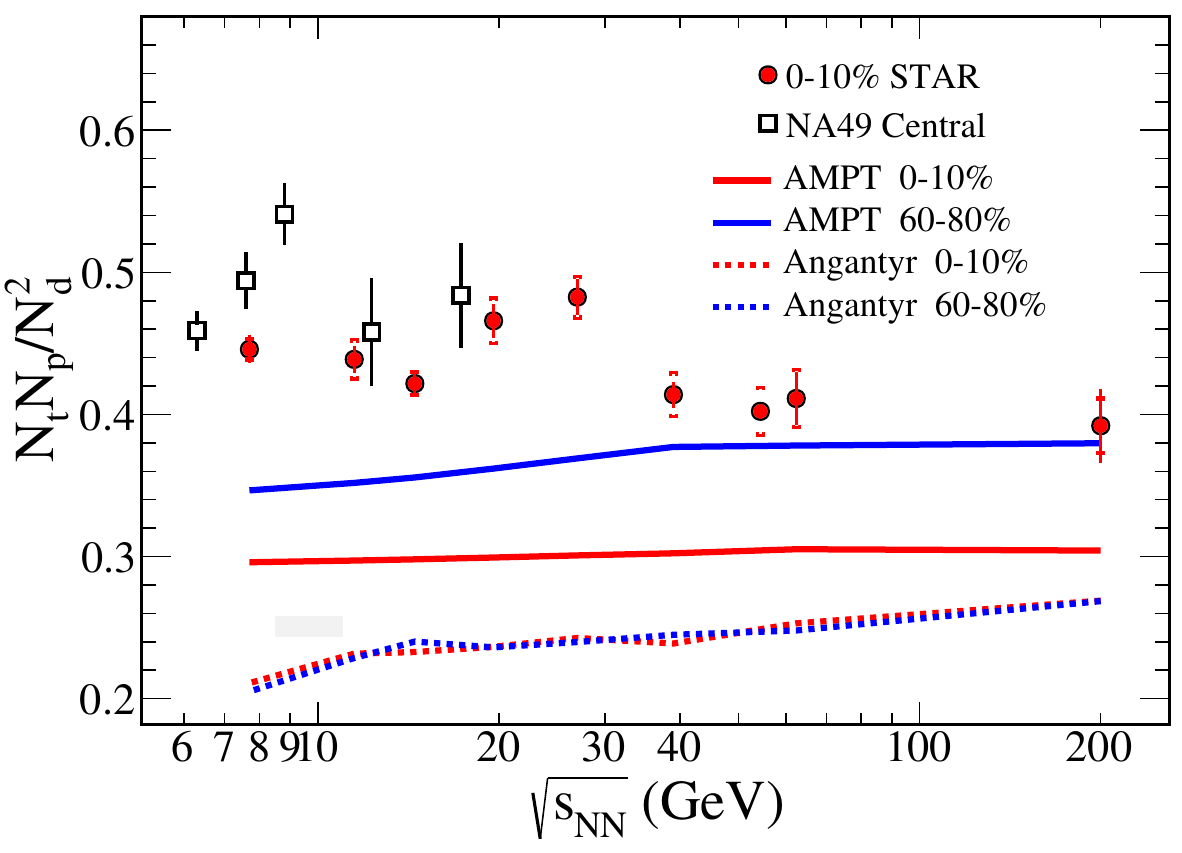}
    \caption{\label{fig6}
     The yield ratio of $N_tN_p/N_d^2$ from the PYTHIA8 Angantyr model, which takes into account the collision energy and centrality, was analyzed in the context of $|y|\leq 0.5$.
      Represented as dash lines from the PYTHIA8 Angantyr model, and as the solid lines from AMPT model~\cite{RN2022}. PYTHIA8 Angantyr model is under the MPI and RC mode.
     Solid circles are the results from STAR detector at 0-10\% central Au+Au collision~\cite{Zhang:2020ewj,2209.08058}.
     Additionally, open squares were used to present the findings from NA49 in central Pb+Pb collision~\cite{RN9,RN19}.
    }
\end{figure}

  Figure~\ref{fig6} presents the experimental results of the STAR detector at 0-10\% central Au+Au collision~\cite{Zhang:2020ewj,2209.08058} , NA49 at central Pb+Pb collision~\cite{RN9,RN19}, then compared with results from the AMPT model and PYTHIA8 Angantyr model. In experimental results, It is evident that there is a non-monotonic energy dependence.
  The yield ratio of light nuclei peaks at $\snn=20\sim30$ GeV, which implies the most significant relative neutron density fluctuations in this energy range.
  Both AMPT model and PYTHIA8 Angantyr model results are lower than
  experimental results; moreover, the AMPT model results which includes QGP medium mechanisms and does not consider critical physics are higher than PYTHIA8 Angantyr model results.
  Due to the absence of critical physics and QGP medium mechanisms, the PYTHIA8 Angantyr model is unable to describe this  non-monotonic energy dependence.

\section{Summary}
\label{sec:summary}

  In summary, by using the PYTHIA8 Angantyr model, we explore the dependence of the relative neutron density fluctuation, neutron-proton correlation $\alpha$, and corresponding light nuclei yield ratio $N_{t}N_{p}/N_{d}^2$ on rapidity, centrality, and collision energy.
  From PYTHIA8 Angantyr model, the relative nucleon density fluctuation cannot be extracted directly from the light nuclei ratio, and the effects of neutron-proton correlation $\alpha$ must be considered in different centralities of collisions.
  The $N_tN_p/N_d^2$ does not change with increasing rapidity coverage and collision centrality, and it increases slightly with increasing collision energy from the PYTHIA8 Angantyr model.
  We show the interplay between multi-parton interactions and color reconnection on the light nuclei yield ratio; there is no effect of CR if MPI is off.
  The experimental results reveal that the light nuclei yield ratio exhibits a peak at $\snn=20\sim30$ GeV, indicating a significant fluctuation in relative neutron density. The non-monotonic energy dependence measurements are underestimated by the PYTHIA8 Angantyr model due to the lack of critical physics and QGP medium mechanisms. Nevertheless, it can be utilized as a baseline in scenarios where critical physics and QGP medium mechanisms are absent.

  This work is supported in part by the Natural Science Foundation of Henan Province (No.212300410386), Key Research
  Projects of Henan Higher Education Institutions (No.20A140024), the
  Scientific Research Foundation of Hubei University of Education for
  Talent Introduction (No. ESRC20220028 and No. ESRC20230002), the
  NSFC 12005114 and NSFC key Grant 12061141008, and key Laboratory of Quark and Lepton Physics (MOE) in Central China Normal
  University (NO. QLPL2022P01).

\end{document}